# Isotropic charge screening of the anisotropic black phosphorus revealed by potassium adatoms

*Zhen Tian,[1,4,5,†] Yu Gan,[2,†] Taiming Zhang,[3] Binbin Wang,[1] Hengxing Ji,[3] Yexin Feng,[2]\* and Jiamin Xue[1]\*.*

\*xuejm@shanghaitech.edu.cn

\*yexinfeng@pku.edu.cn

[1]School of Physical Science and Technology, ShanghaiTech University, Shanghai 201210, China

[2]School of Physics and Electronics, Hunan University, Changsha 410082, China

[3]School of Chemistry and Materials Science, University of Science and Technology of China, Hefei 230026, China.

[4]Shanghai Institute of Optics and Fine Mechanics, Chinese Academy of Sciences, Shanghai 201800, China

[5]University of Chinese Academy of Sciences, Beijing 100049, China

[†]These authors contributed equally to this work

**Abstract**

**Black phosphorus has attracted great research interest due to its numerous applications in electronic devices, optoelectronic devices, energy storages and so on. Compared with the majority of two-dimensional materials, black phosphorus possesses a unique property, i.e. the strong in-plane anisotropy. All the properties reported so far, including its effective mass, electron mobility, light absorption, thermal conductivity and so on, have shown great anisotropy in the basal plane. This property renders black phosphorus unique applications not achievable with other two-dimensional materials. In this work, however, we discover a remarkable isotropic behavior in the strongly anisotropic black phosphorus, i.e. its electrostatic screening of point charges. We use the tip-induced band bending of a scanning tunneling microscope to map out the Coulomb field of ionized potassium adatoms on black phosphorus, and reveal its isotropic charge screening. This discovery is important for understanding electron scattering and transport in black phosphorus.**

**I. INTRODUCTION**

With its layered structure, high charge-carrier mobility [1-8], highly tunable layer-number dependent band gaps [6-10] and so on, black phosphorus (BP) are attracting a large amount of research efforts. Compared with most of the other 2D materials, BP has an intriguing in-plane anisotropy due to its puckered lattice structure in the basal plane, which gives BP many unique properties and applications, such as the one-order-of-magnitude different effective masses along its zigzag and armchair directions, the anisotropic charge carrier transport [5-8], polarization dependent light absorption [5,8,10,11], anisotropic thermal transport [12-15] and so on. To the best of our knowledge, all the measured physical properties of BP reported so far show strong anisotropy in the basal plane.

In this work, a remarkable isotropic property of the anisotropic BP is discovered, i.e. the electrostatic screening of point charges. Screening of ionized impurities greatly affects the charge transport in BP based electronic devices, especially at low temperatures. To probe the screening of point charges at the atomic scale, we first create charged impurities, i.e. ionized potassium (K) adatoms, on top of the BP surface. Then we utilize the effect of tip-induced band bending (TIBB) [16,17] of a scanning tunneling microscope (STM) to control the charge states of the adatoms. The TIBB coupling between the STM tip and the ionized adatoms is mediated by the screened Coulomb field, and the spatial distribution of the coupling unambiguously reveals the isotropic screening of point charges in BP.

The charge screening in BP has been studied by several theoretical groups [18-21]. However, very contradicting results have been predicted regarding whether the screening is isotropic or anisotropic. Our study is the first experimental effort to solve this puzzle. The analysis employed here can also be adopted to probe the screening behavior in other related materials with predicted strong dielectric anisotropy, such as SnS, SnSe and so on [22]. Previously, only macroscopic methods can be used to measure the dielectric anisotropy [23,24]. Our study could provide a new route for probing this behavior at the atomic level, which is more relevant to electron interaction and transport in these materials.

II. RESULTS

The bulk BP crystals were grown by a low-pressure transport method (see the Supplemental Materials, SM, for details). The bulk BP was exfoliated in ultrahigh vacuum (< $10^{-10}$ mbar) and then cooled down to 4.5 K for the STM and STS measurements. Fig. 1a shows the typical topography of a pristine BP surface, where the atomic chains (zigzag direction) can be clearly seen. The armchair direction is perpendicular to the atomic chains and is marked with the red dotted arrow in Fig. 1a. The dumbbell-shaped features in the topography come from standing waves of electrons scattered by its intrinsic vacancies, which has been discussed previously [25,26]. Their highly anisotropic structure stems from the underlying electron dispersion relation. Since the Fermi surface is strongly elongated along the zigzag direction, the standing wave has a much shorter wavelength along the atomic chains.

To study the screening of K adatoms by BP, we evaporated K onto the pristine BP surface inside the STM chamber with a getter source (see SM for details). Fig. 1b shows the topography after K doping. In order to avoid moving the adatoms unintentionally, a large tip-sample separation (2 pA tunneling current at − 0.7 V sample bias) is used to obtain the image. We found that K adatoms can be easily dragged by the tip along the zigzag but not the armchair direction (see Fig. S1d in the SM), presumably due to a much higher diffusion energy barrier along the armchair direction. To confirm this hypothesis, density functional theory (DFT) calculations were performed on the diffusions of K adatoms along the two directions with the climbing image nudged elastic band method (cNEB) [27]. We find that the hollow site is the most stable adsorption site for K adatom (Fig. S2) and the diffusion energy barrier (Fig. S3) along the armchair direction (~ 0.28 eV) is one order of magnitude higher than that along the zigzag direction (~ 0.025 eV). This result adds yet another interesting anisotropic behavior to BP and is directly related to ion mobility in BP, which is an important factor in BP based energy storage applications [28,29].

Due to the low tunneling current, atomic resolution cannot be achieved in Fig. 1b, but the feature of an intrinsic vacancy (inside the blue dotted box) points out the crystal directions and the zoom-in images in Fig. S1 confirm that. The K adatoms seen in Fig. 1b have lateral radius of ~ 2 nm and dopant density about $1 \times 10^{12}$ cm$^{-2}$. When we scan the same area with higher sample bias, almost every adatom will be seen surrounded by a disk-like feature (Fig. 1c). As we will discuss in details later, this feature is associated with putting electrons back onto the positively charged K nuclei. Thus it indicates that almost all the adatoms are ionized at 4.5 K. Recent angle-resolved photoemission spectroscopy (ARPES) study [30] showed that due to the strong electric field generated by the K nuclei, the band gap of BP can be greatly modulated by the Stark effect. We then perform STS on BP with K doping and compare it with that on pristine BP (Fig. 1d). With this doping level, we can only see Fermi level shift away from the valence band by ~ 100 meV, without significant change of its band gap. In the ARPES study the K coverage for closing the band gap is about 0.36 monolayer, which is far higher than the density shown in Fig. 1b. Our DFT calculations also show good agreements with the experimental results. The change of band gap at different K dopant densities can be found in Fig. S4 and S5, which show that the band inversion occurs at about the K dopant density of $1.65 \times 10^{14}$ cm$^{-2}$, two orders of magnitude higher than that in our experiment.

Fig. 1c shows interesting disk-like features. We focused on one of them and get high resolution topography (Fig. 2a) together with its d$I$/d$V$ image (Fig. 2b). In Fig. 2a, the atomic corrugation of underlying BP is not disturbed by this feature. It indicates that the apparently higher part inside the disk is not of structural but of local density of states (LDOS) origin. However, in Fig. 2b, only the edge of this disk has shown up as a bright ring with no enhancement of LDOS inside the ring. The bright and dark oval shapes at the center can be explained by the Friedel oscillation in BP caused by the adatoms [25] (see Fig. S10 and related discussion). Another character of the disk-like features is that the radius of the disk will decrease when the magnitude of the sample bias is reduced, which can be found in the d$I$/d$V$ mapping videos and Fig. S6d, S6e and S6f in the SM. All these behaviors can be understood based on the theory of TIBB [16,17], which has been studied extensively in Si-doped [31,32] or Zn (Mn) -doped [33-35] GaAs systems, Co-doped graphene system [36,37], ZnO system and so on [38-40]. In these systems, when the negatively biased tip is scanning around, it will move up the energy levels of dopants (for GaAs, ZnO and so on) or adatoms (for graphene) due to TIBB. When the TIBB is large enough to shift the dopant energy levels above the sample Fermi level, the originally neutral dopant or adatoms will be ionized (charged). This results in a disk-like feature in the topography or a ring in the d$I$/d$V$ map around a dopant or an adatom.

With apparent similarities to the previously studied systems, however, the tip-induced charging process in K-doped BP has its unique properties. Opposite to those systems, the disk or ring features only present at negative sample bias but not the positive side (see Fig. S1c and also the videos). This can only be explained if the K adatoms are already ionized at 4.5 K, and the ring around each adatom is a signature of putting one electron back onto the nuclei instead of ionizing it. With this understanding, we draw the schematics in Fig. 2c. When the tip is laterally far away from an adatom (i.e. $d_{xy} \gg r$, where $d_{xy}$ is the lateral distance between the tip and adatom; $r$ is the radius of the disk-like feature seen in topography), TIBB will not influence the band near the adatom (Fig. 2c left) and the adatom is already ionized. Due to its Coulomb potential, the band around it is pulled down in energy. When the tip is near the adatom ($d_{xy} \leqslant r$), even without a bias voltage, the band will be bent due to the work function difference between the tip (4.0~ 4.5 eV) [33] and sample (4.5 eV ~ 5.0 eV) [41] (Fig. 2c middle). If a negative sample bias is applied, the TIBB is further increased. When the K adatom level is beneath the sample Fermi energy (blue dotted line), the K nuclei become neutral and the Coulomb potential induced band bending of the ionized K (the red dotted line in Fig. 2c right) vanishes. The amount of states in the valence band available for tunneling to the tip will be enhanced, which results in an instantaneous increase in tunneling current and make the tip retract to maintain the constant current. So there is a height increase in the constant-current topography as shown in Fig. 1c and 2a. While for the differential conductance map in Fig. 2b, the tip with 0.7 V is just enough to trigger this charging event when it is located at distance $r$ away from the adatom. With the tip inside this ring the adatom is already in its neutral state, so the differential conductance is similar to the values obtained when the tip is far away from the adatom.

With this understanding of TIBB in BP, we further analyze the ring and extract quantitative information about the electrostatic screening. As shown in Fig. 3a (Fig. S7a shows corresponding topography for higher bias), the TIBB produces an almost perfect circle around an isolated K cluster. Since the TIBB pattern is the result of the Coulomb potential of the tip and charged adatoms screened by BP electrons, the circular shape in Figure 3a unambiguously reveals its isotropic behavior. To get more quantitative analysis, we performed line spectroscopies along the armchair (Fig. 3b) and zigzag (Fig. 3c) directions indicated by the dotted arrows in Fig. 3a. Two hyperbola-like features are clearly visible in both the line spectroscopies. The outer hyperbola-like feature corresponds to the first charging event induced by TIBB, and the inner one corresponds to a second charging event after one electron is already put onto the K cluster. This phenomenon indicates that the cluster contains more than one K nucleus. Comparing Fig. 3b and 3c, we can see that the TIBB effects are almost identical along the two crystal directions. We further calculate the TIBB based on the model introduced by Feenstra [16,17] and show the results in Fig. 3d. Each colored curve is the calculated TIBB as a function of the lateral distance between the tip and the K adatom at a certain sample bias. Although many parameters were used in the model calculation, most of the parameters can be found in the literatures about BP [2,18,41] (the parameters we used in the calculation are shown in Table S1 in the SM).

The calculated data in Fig. 3d adopted a work-function difference of – 0.5 eV and a tip-sample distance of 0.5 nm. The grey horizontal line marks the TIBB that best matches with the data in Fig. 3b and 3c. The calculated data points fall nicely on top of the measured hyperbola-like curves. We can see that the K energy level is ~ 230 meV above the Fermi level, slightly (~ 30 meV) above the conduction band edge of BP. The quantitative results corroborate the schematics in Fig. 2c. Notice that the reason we can use the calculation with one set of parameters to fit the data along the two crystal directions roots in the isotropic screening behavior of the electrons in BP. If the screening were anisotropic, we would get different TIBB along the armchair and zigzag directions, which is absent in our experiments. Some calculations [18] predicted a smaller electrostatic dielectric constant (~ 10.2) along the zigzag direction compared with that of the armchair direction (~ 12.5) (another DFT calculation predicted similar anisotropy [20]). In Figure 3d, we plot with dotted line a calculated TIBB with a 20% smaller dielectric constant. Due to the reduced screening, to get the same 230 meV band bending to charge the K nuclei, the tip need to be ~ 1 nm further away from the center. This will result in a strongly elliptical shape of the ring around the adatoms, elongated along the zigzag direction. The calculated charging ring with theoretical predicted anisotropic dielectric constant tensor is superimposed on Figure 3a (see Figure S8 for calculation details), which shows clear difference from the measurement. We also tested different parameters in the simulation, which all show the same anisotropy (Table S2). Recent simulations [19,21] using effective Hamiltonian with the anisotropic electron dispersion relation showed that the screening should be isotropic for long range potentials caused by point charges. However, the influences from high-energy bands and ionic contributions are not included. Our results would stimulate further theoretical efforts to investigate the underlying mechanisms of the isotropic screening of anisotropic BP.

So far, we have focused on individual K adatoms. Next we study the interaction between them. Fig. 4a shows such a case with four K adatom clusters. Each cluster has its own ring structure in the d$I$/d$V$ map (Fig. 4b). Due to the closed spacing between the adatoms, the rings intersect with each other. Surprisingly, at the intersecting points the rings show anti-crossing. Examining Fig. 4b closely, we can see that the ring diameters shrink after the intersecting points. Clusters 1 and 2 are closer to each other than 3 and 4, and the shrinking is stronger for the rings coming from 1 and 2 than that from 3 and 4. Combining these observations with the previous discussion of TIBB of individual adatom, we can understand this behavior as a result of donor-donor interaction. As shown schematically in the lower part of Fig. 4d, when the tip charges the right one of the two closely spaced K clusters ($D_1$ and $D_2$), it raises up the energy level of $D_1$, hence more TIBB is required to charge the left cluster. From the calculation in Fig. 3d, a larger TIBB corresponds to closer tip-adatom distance at a certain sample bias. This explains the shrinking ring diameter after the intersection. To quantitatively determine how much shrinking that has happened and its relation with the charge screening property of BP, we measured an area with only one pair of intersecting rings as shown in Fig. 4c (its corresponding large scale topography is shown in Fig. S9b). The dotted lines are parts of circles used to identify the edges of the disks. For clarity, the upper part of Fig. 4d shows lines extracted from the − 0.9 V bias topography. The radius $r$ of a ring before intersecting and its change $\Delta r$ after intersecting can be measured. From these we obtain the change in TIBB ($\Delta\varphi$) from the curves in Fig. 3d. $\Delta\varphi$ can also be calculated as $\Delta\varphi = e^2/(4\pi\varepsilon_0\varepsilon_r d)$, where $e$ is the electron charge, $\varepsilon_0$ is the vacuum permittivity, $d$ is the distance between the adatoms, and $\varepsilon_r = (\varepsilon_{BP} + 1)/2$ is the modified dielectric constant since the K adatoms are located at the interface between BP and vacuum. Using this formula and the data extracted from Fig. 4c we calculate the dielectric constant of BP $\varepsilon_{Bp}$ to be 13. This value is close to the one used in the TIBB calculations for Fig. 3, demonstrating the self-consistency of our analysis.

## III. DISCUSSION

Interestingly, the interaction between closely located adatoms can also be interpreted based on a coupled double-quantum-dot system. The K adatoms can be viewed as quantum dots while the tip behaves like a scanning gate to tune the energy levels in each quantum dot. We can use (0, 0) to denote the condition when both quantum dots (i.e. adatoms) are positively charged, (1, 0) to denote when one electron is put onto the left quantum dot and so on. Then the data in Figure 4 can be viewed as the stability diagram [42] for coupled quantum dots as shown in Figure S11 of the SI. When the tip crosses the borderline between two different stability regions (e.g. from (0, 0) to (1, 0) or (0, 1)), one electron is added on one of the quantum dots. The most interesting areas are the regions between (1, 0) and (0, 1), highlighted with the dotted circles in Figure S11, where there is no borderline since the total electron number in the couple system does not change when the tip moves from one region to the other. Similar pattern has been observed in scanning gate microscopy of coupled carbon nanotube quantum dots. [43]

## IV. CONCLUSION

In conclusion, we have used K adatoms as probes to study the electrostatic screening of point charges in BP. This was achieved by quantitatively analyzing the charging rings around individual and coupled K adatoms induced by the TIBB effect of an STM. We revealed its isotropic screening behavior despite the underling strongly anisotropic electron dispersion. To better appreciate the uniqueness of this property in BP, we have listed related anisotropic properties of BP in Table S3 of the SM. All the listed properties except for the charge screening in BP are anisotropic. This finding is of great importance for us to understand charge carrier transport and scattering in BP, which will benefit its device applications. The analysis used in this study can also be employed to probe the screening behavior in other related materials and study the more detailed relationship between screening and local environment at the atomic level.

**Acknowledgments**

Z.T., B.W. and J.X. are supported by the Ministry of Science and Technology of China (No. 2017YFA0305400), Thousand Talents Program and ShanghaiTech University. Y.G. and Y.F. are supported by the National Natural Science Foundation of China (Nos. 11604092 and 11634001) and the National Basic Research Programs of China (No. 2016YFA0300900). The computational resources were provided by the computation platform of National Supper-Computer Center in Changsha, China. T.Z. and H.J. are supported by National Natural Science Foundation of China (No. 51672262, 51761145046) and 100-Talent Program of Chinese Academy of Science.

**Figures and Captions**

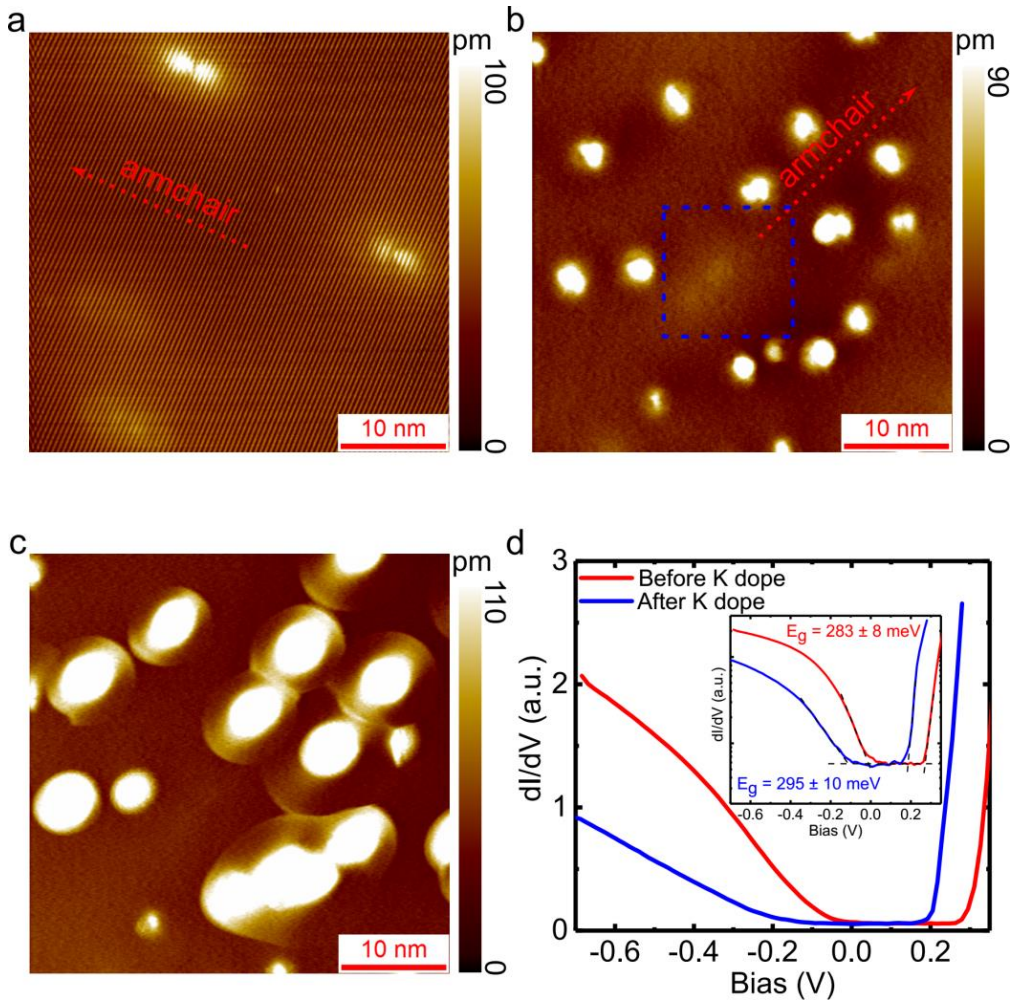

**FIG 1. (a)** Topography of BP before K doping (scan parameter: – 0.5 V, 100 pA). **(b, c)** Topographies after K doping (scan parameters: – 0.7 V, 2 pA for **b** and – 1.4 V, 2 pA for **c**). The red dotted arrows indicate the armchair directions. **(d)** Comparison of d$I$-d$V$ spectra between pristine BP and K-doped BP. STS parameters: lock-in frequency $f$ = 991.2 Hz, modulation voltage $V_{rms}$ = 10 mV, and $I$ = 50 pA. The inset is the data plotted in semi-log scale for a better determination of the band gaps.

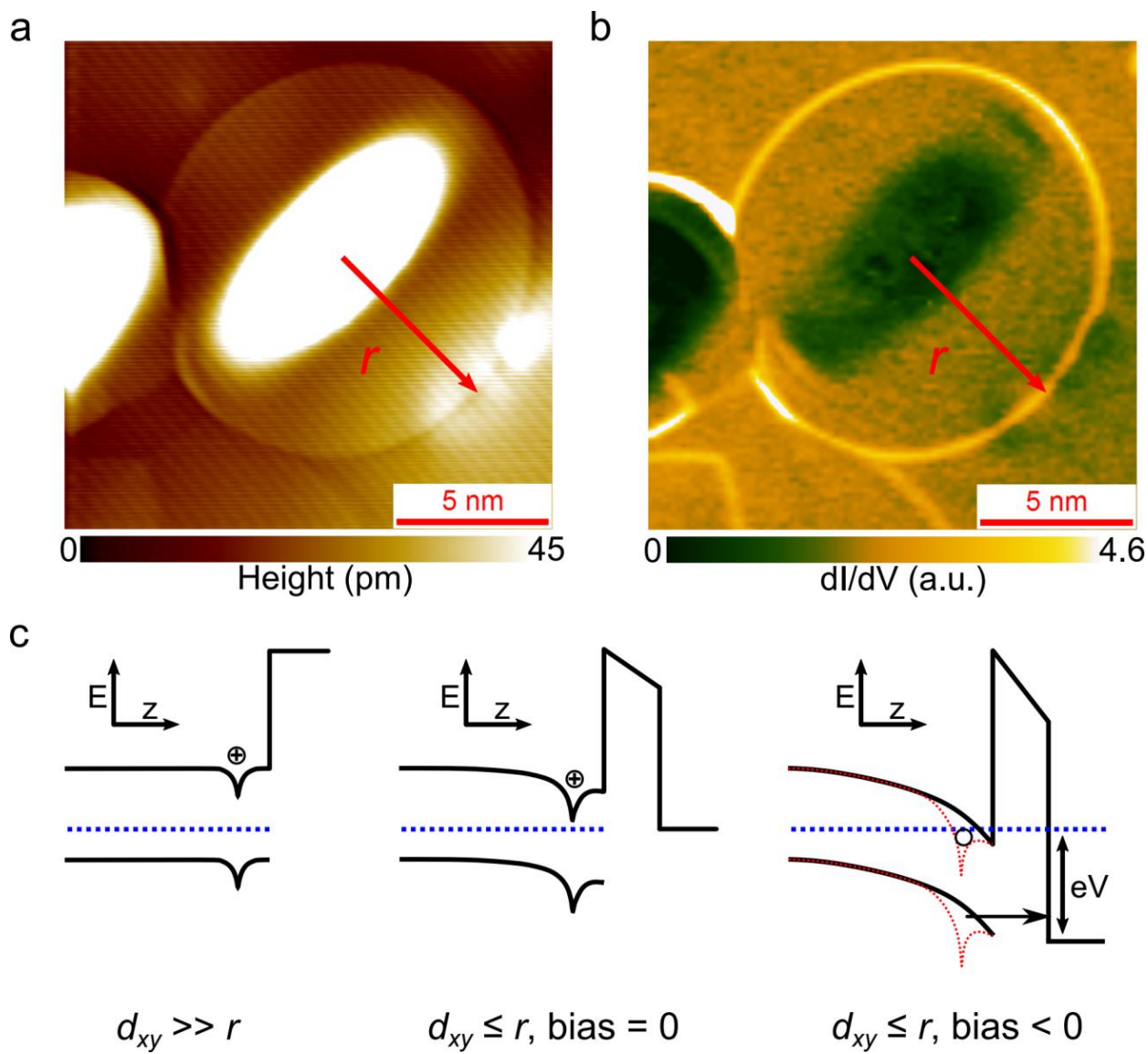

**FIG 2.** (**a**) High resolution topography (− 0.7 V, 50 pA). (**b**) Corresponding differential conductance map (− 0.7 V, 50 pA). (**c**) Schematics of the charging mechanism of K adatoms. Left: tip is laterally far away from the adatom; middle: tip is near the adatom with zero sample bias; right: tip position is the same as that in the middle, but with negative sample bias. Black solid lines represent band edges and red dotted line the band edges before charging. Blue dotted lines show the Fermi level of the sample.

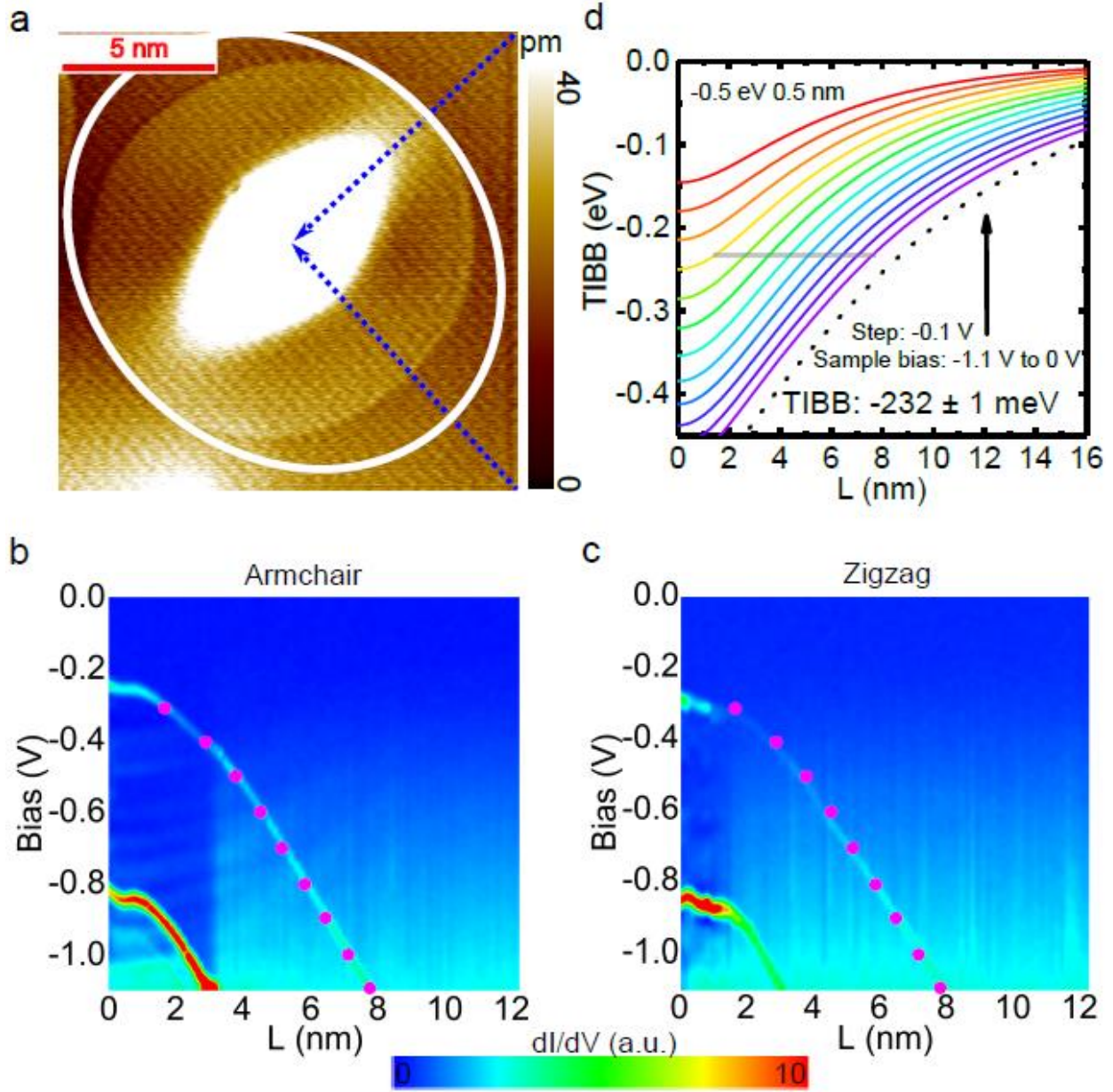

**FIG 3.** (**a**) Topography on an isolated K cluster (− 0.7 V, 10 pA). The white ellipse outlines the calculated TIBB charging disk, assuming the anisotropic dielectric tensor predicted by DFT calculations. See Figure S8 in the SI for more details. (**b,c**) The d$I$-d$V$ spectra along the blue dotted arows in **a**. STS parameters: lock-in frequency $f$ = 991.2 Hz, modulation voltage V$_{rms}$ = 10 mV, and $V_{initial}$ = − 1.1 V, $I_{initial}$ = 100 pA. The circles are calculated data points taken from **d**. They are the intersection points between the grey horizontal line and the colored TIBB curves.

(**d**) Calculated TIBB as a function of the tip-adatom lateral distance at different sample biases with an isotropic dielectric constant of 12.9. The dotted line is a calculated TIBB at sample bias – 1.1 eV with 20% smaller dielectric constant compared with that of the colored lines. The gray horizontal line intersects with the calculated TIBB curves to give the data points in (b) and (c).

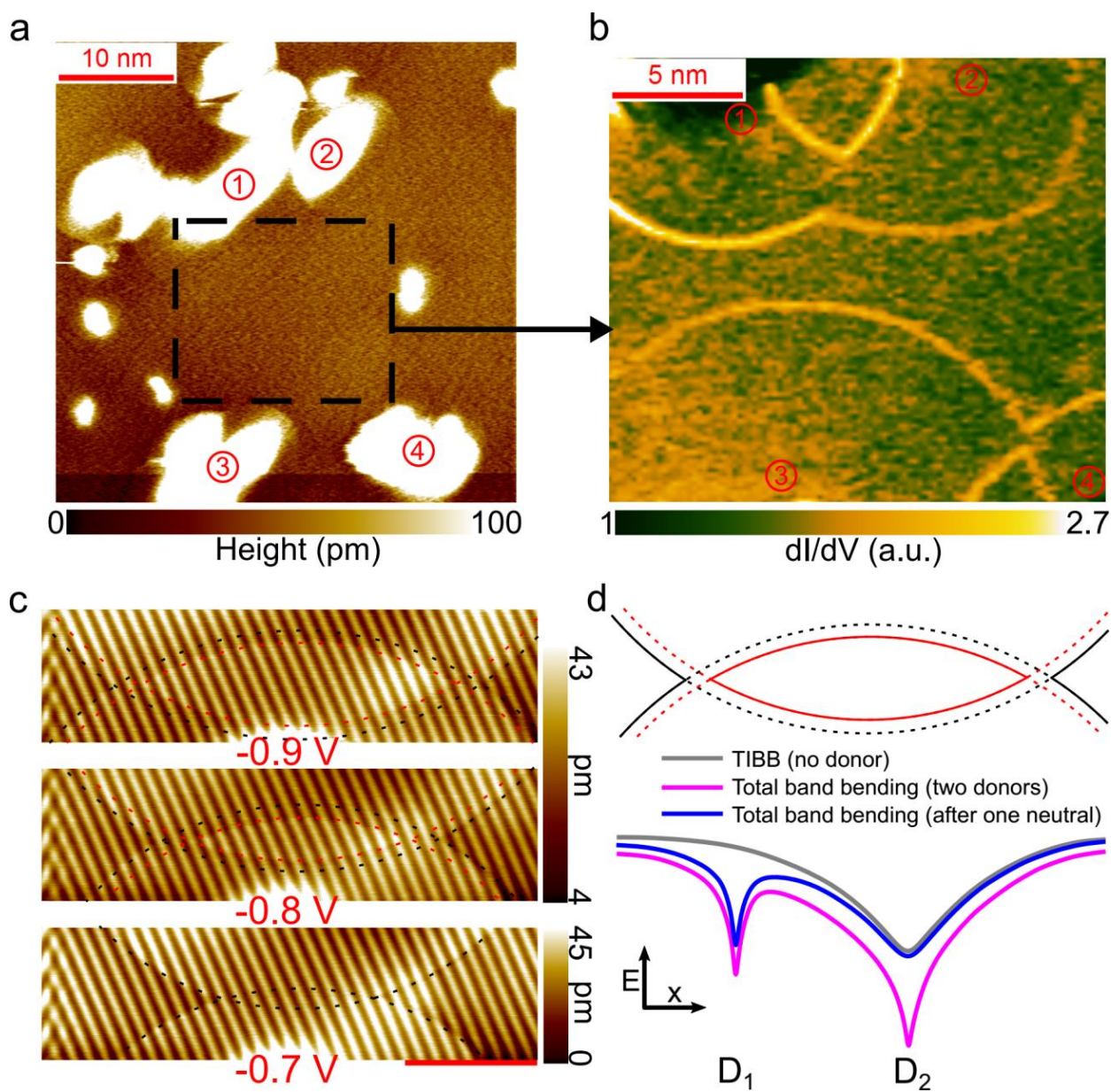

**FIG 4.** (**a**) Large area topography with four K adatoms ($V = -1$ V, $I = 1$ pA). (**b**) Differential conductance of the area inside the black dashed box in **a** with $V = -0.8$ V, $I = 40$ pA. (**c**) High resolution topography of one pair of intersecting rings at tunneling current 50 pA and different sample bias. The black and red dotted lines identify the edge of outer and inner disks. Scale bar = 4 nm. (**d**) Upper part shows the extracted lines from the $-0.9$ V bias topography (**c** upper panel) with solid lines representing the visible disk edges in the image and dotted lines the extension of solid lines; lower part shows the schematic representation of donor-donor interaction with tip above the right K adatom, $D_2$.